\documentclass[mathleft]{an}
\usepackage{graphicx}
\usepackage{times}
\newcommand{\dd}{$\rm deg^{2}$}
\newcommand{\flux}{$\rm erg \, s^{-1} \, cm^{2}$}

\begin{document}

% The following seven commands are intended for editorial usage and should be ignored by
% the author(s).
	%\Pagespan{789}{}% Document's page range. 
% If second parameter is left empty, the last page is computed automatically.
	%\Yearpublication{2006}%
	%\Yearsubmission{2005}%
	%\Month{11}%   
	%\Volume{999}%  
	%\Issue{88}% 
% \DOI{This.is/not.aDOI}% 

\title{The XMM-LSS cluster sample
        and its cosmological applications.\\
        Prospects for the XMM next decade}
        
\author{M. Pierre\inst{1}\fnmsep\thanks{Corresponding author:
  \email{mpierre@cea.fr}\newline}
%Example 
%for footnote, note the usage of the \texttt{fnmsep}
%command as separator between institute number and footnote mark} 
\and  F. Pacaud\inst{1,2}
\and J.B. Melin\inst{3}
\and the XMM-LSS consortium}
\titlerunning{The XMM-LSS survey}
\authorrunning{M. Pierre}
\institute{
DAPNIA/Service d'Astrophysique,  Laboratoire AIM  CNRS, CEA-Saclay, F-91191 Gif-sur- Yvette, France
\and 
Argelander-Institut f\"ur Astronomie, University of Bonn, Auf dem H\"ugel 71, 53121 Bonn, Germany 
\and 
 DAPNIA/Service de Physique des Particules, CEA Saclay, F-91191 Gif-sur-Yvette, France.  }

\received{: September 2007}
\accepted{: November 2007}
%\publonline{later}


\abstract{%
The well defined selection function of the XMM-LSS survey enables a simultaneous modelling of the observed cluster number counts and of the evolution of the L-T relation. We present results pertaining to the first 5  \dd\ for a well controlled sample comprising 30 objects : they are compatible with  the WMAP3 parameter set along with cluster self-similar evolution.  Extending such a survey to 200 \dd\  would (1) allow discriminating between the major scenarios of the cluster L-T  evolution and (2) provide a unique self-sufficient determination of $\sigma_{8}$ and $\Gamma$ with an accuracy of  $\sim$ 5\% and 10\% respectively,  when adding mass information from weak lensing and S-Z observations.}
\maketitle

\section{Introduction}

It has been recognised for a long time that clusters of galaxies, as the most massive entities of the universe, can be used as cosmological probes. They provide key information on the normalisation of the power spectrum and are potentially suitable for studying the properties of dark energy. They represent important and independent constraints in addition to those from  the CMB and supernovae because they  involve  very different physics. It is also critical to ensure consistency between the cosmological constrains from the early and local universe. 

Main statistical tools for cluster studies are the cluster number counts ($dn/dz$) and the cluster two-point correlation function ($\xi$).   This requires that, whatever the detection wavelength,    the samples must be in some sense, complete and uncontaminated and thus, requires well understood detection and selection procedures. Cluster physics evolution is a key  ingredient in interpreting the observed cluster density as function of redshift. It is usually modelled in the form of scaling laws relating  observable quantities such as flux,  richness, luminosity or  temperature to cluster masses. The  cluster scaling laws are, however, still poorly known beyond the local universe.

With its mosaic of overlapping  XMM pointings ($10^{4}$ s), the XMM 
Large-Scale Structure survey (XMM-LSS, Pierre et al 2004)  has been designed to detect a significant fraction of the cluster population out to $ z 
= 1$, over an area of several tens of \dd , so as to constitute a sample suitable for cosmological studies. We present below the procedures developed to detect the clusters and to further analyse  their number counts along with their temperature and luminosity distribution in a self-consistent approach. In light of the results obtained so far, we discuss the cosmological impact of a future 200 \dd\ XMM wide survey. 

\section{Detecting and selecting clusters in the XMM-LSS survey}

 In the redshift range of interest, although the cluster apparent sizes ($20'' < R_{c} < 100 ''$) are significantly larger than the XMM PSF and source confusion can be considered as negligible, cluster detection is a very specific task since our objects  are weak sources (count-rate from 0.3 to 3 counts/min). We developed a two-step procedure combining wavelet multi-resolution analysis and maximum likelihood fits both using Poisson statistics. The pipeline was extensively tested using simulations which allowed us to define a sub-region in the {\tt extent} vs {\tt extent likelihood} parameter space, where the contamination level by  point-sources is lower than 1\%. This constitutes the {\em class one} (C1) cluster sample. Strictly speaking, this selection is not flux limited, but allows the construction of well controlled and uncontaminated cluster samples significantly larger than a simple flux limit would allow (Pacaud et 2006). Our C1 sample shows a density of $\sim 6$ clusters per \dd\ .

\section{Current cluster results}
\begin{figure}
\hspace{-1.5cm} \includegraphics[width= 9.8cm]{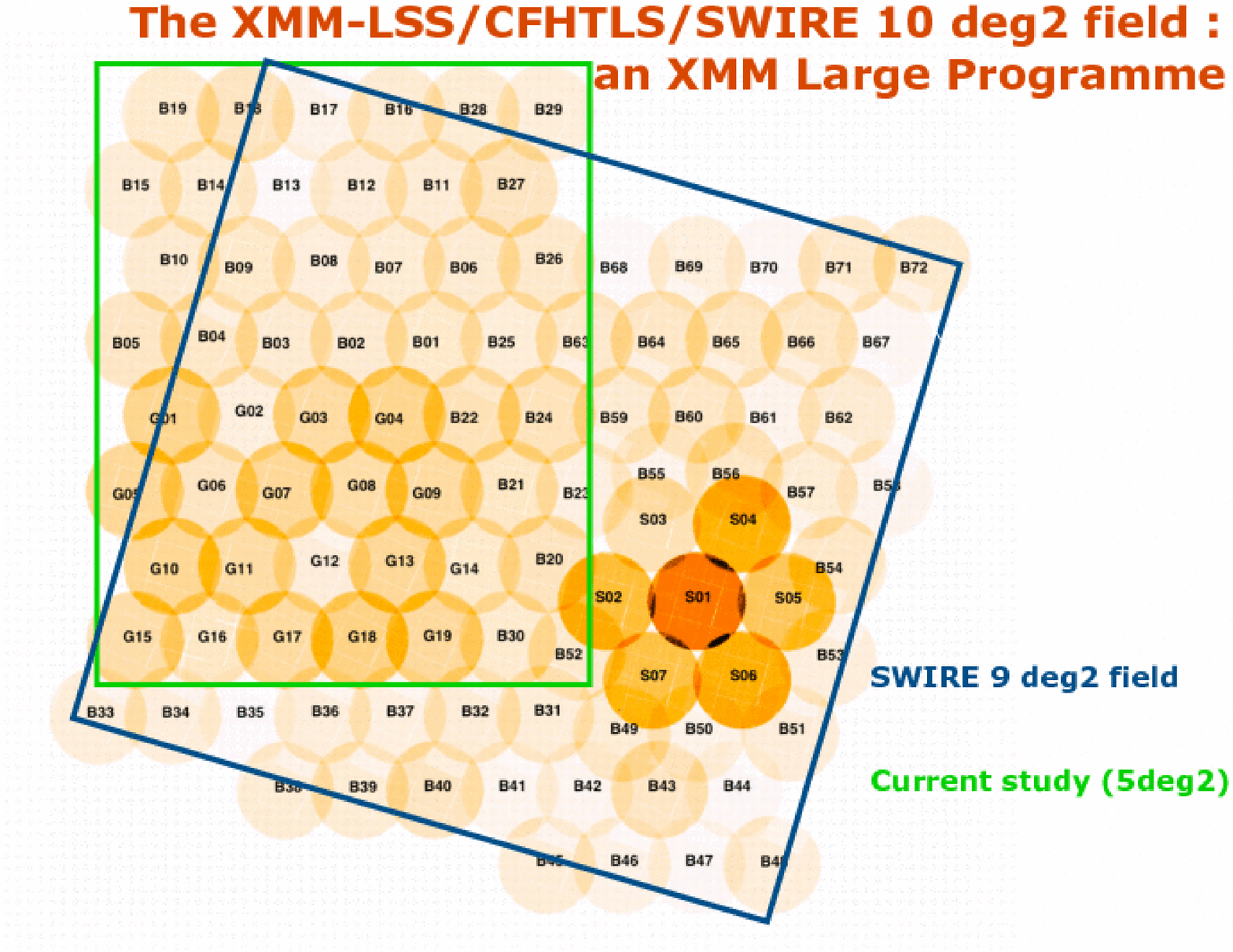}
\caption{{\bf Layout of the 98 XMM-LSS pointings} including the Subaru Deep Survey. The colour scale indicates the effective exposures, from 80 ks (SDS01) to $\sim$ 0.  The green rectangle delineates data obtained prior to the AO5. In addition to the  coverage by the SWIRE  and CFHT Legacy Surveys, observations from the VLA, Integral, UKIDSS and Galex are available in the field. Full coverage by SCUBA2 and Herschel is planned. }
\label{map}
\end{figure}

The XMM-LSS currently covers 10 \dd. It is located in the W1 area of the Canada-France-Hawaii Telescope Legacy Survey (CFHTLS)  and associated with a number of surveys in the radio, infrared and UV domains (Fig.  \ref{map}). In these proceedings, we summarise the results from the first 5 \dd\,   which are presented in detail by Pacaud et al 2007 (P07).

\subsection{Modelling the cluster number counts} 
\begin{figure}
\includegraphics[width= 8.3 cm]{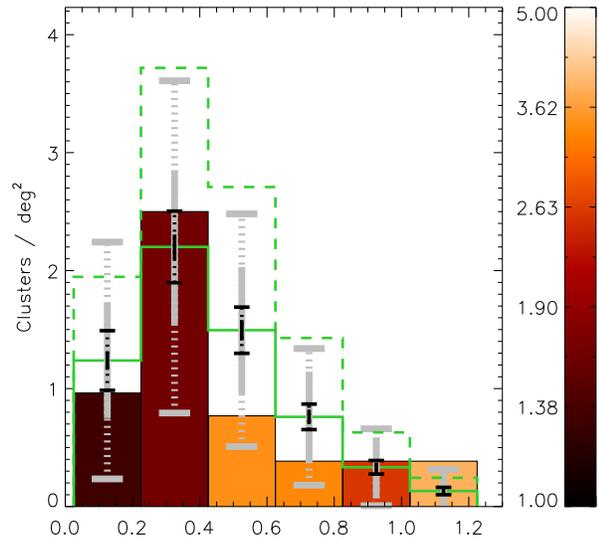}
\caption{{\bf The current C1 cluster redshift distribution over the first 5 \dd\  of the XMM-LSS}. The colour  scale indicates the cluster mean temperature for each bin (unweighted mean of the individual cluster temperatures in keV.)  The solid green histogram shows 
the expectations of our cosmological model (WMAP3: $\sigma_{8}$ = 0.74 and self- 
similar evolution for the Lx-T relation) along with the Poisson error bars. 
The dash-line histogram shows the expectations for a model with WMAP 1st year 
cosmological parameters (  $\sigma_{8}$  = 0.85) and a non-evolving Lx-T relation. 
Fluctuations around the mean expectation are represented by the solid and dotted error bars for the shot noise and sample variance (estimated from Hu \& Kravtsov  2003) respectively. The grey error bars are for 5 \dd, while the black ones for 200 \dd.  }
\label{dndz}
\end{figure}

Some 30 C1 clusters are found in the first 5 \dd . They all have been spectroscopically confirmed. Their redshift distribution is displayed on Fig. \ref{dndz}. P07 performed an ab initio modelling of the observed number counts as follows: assume a (1)  cosmological framework ($\Lambda $CDM)  and a  power spectrum along with a transfer function; (2) a mass function, (3) a halo model, (4) various scaling evolutionary relations for cluster physics. Then, for each redshift and mass range, the predicted luminosities are transformed into XMM count-rates using a dedicated plasma code. The C1 selection criteria are finally applied, resulting in a simulated observed redshift distribution  (Fig. \ref{dndz}). The effect of the degeneracy between cluster scaling law evolution and cosmology is clearly illustrated on the figure;  however, our results, despite the still small size of the sample, favour the WMAP3 parameter sets along with cluster self-similar evolution.
 
\subsection{Evolution of the L-T relation}
 
 Each C1 cluster undergoes dedicated spatial and spectral fits in order to derive reliable luminosity and temperature estimates. Willis et al (2005) have shown that using a well adapted binning procedure, it is statistically possible to obtained a 20\% temperature accuracy with only 200 photons for groups up to 2 keV. It turns out that for all C1 clusters,  we obtain  temperature measurements with satisfactory accuracy. 
The average temperature of  the C1 clusters as a function of redshift is displayed on Fig. \ref{dndz}. Because of the very tight relation  between X-ray temperature and luminosity, the mean temperature of the detected clusters appears to increases with redshift (Malmquist bias).  This redshift - temperature distribution shows that the XMM-LSS survey unveils for the first time the population of low-mass groups (T= 2 keV) around $z=0.3$, which constitute the building blocks of the present day clusters.

\begin{figure}
\includegraphics[width= 8.3 cm]{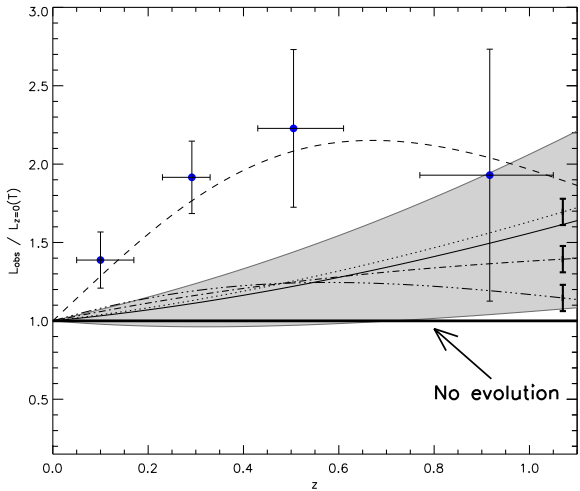}
\caption{{\bf The L-T relation}. The graph shows the cluster X-ray luminosity enhancement with respect to the expectation at $z=0$. Our clusters are sorted in 4 redshift intervals. The dash line is the result of an ad hoc two-parameter fit to the raw data points: $(1+z)^{\alpha} \times E(z)^{\beta}$ with  $\alpha = 4.7, ~ \beta = -5.4$. When the survey selection function is taken into account, the best fit is the solid line, $ (1+z)^{\alpha}E(z)$  with $\alpha = - 0.07$, which is very close to the evolution predicted by the self-similar model (dotted line, $\alpha = 0$ ). The grey region delineates  the 1$\sigma$ confidence interval for the surveyed 5 \dd. The dot-dash and triple-dot-dash lines are the  evolutionary models by Voit (2005) including non gravitational physics. The thick error bars at the end of the three models indicate the expectations for a 200 \dd\ survey.  }
\label{LT}
\end{figure}

Assuming that all clusters, whatever their mass, follow the same evolutionary scaling laws, we have used our data to constrain the evolution of the L-T relation. The observed luminosity enhancement, with respect to the local expectation, is computed   for the 30 clusters distributed in four redshift bins (Fig \ref{LT}). The raw data suggest a rather strong evolution, best fitted with a two-parametre model. However, the inclusion of the survey selection function in the fit suggests a much more mild evolution, fully compatible with the predictions of the self-similar model.  The  fact that selection effects were not systematically considered in the former L-T(z) studies may explain the discordant results obtained so far (see a compilation in P07): because the cluster mass function is so steep, most clusters are detected around the survey limiting sensitivity, hence favouring the compilation of over-luminous objects at any redshift. This is certainly not a feature unique to the XMM-LSS survey, but rather of any X-ray cluster survey (provided that the detection is performed down to the  capabilities of the survey). This bias is to affect any  sub-sample ``randomly'' selected from all-sky or serendipitous surveys for subsequent deep XMM or Chandra temperature observations. We emphasise that the XMM-LSS survey represents the first attempt, not  only to determine the L-T  relation with survey data alone but also to explicitly  include the selection effects in the determination of the evolution of the scaling laws. The latter was possible thanks to the very well modelled survey selection function.

 \section{AGNs in the XMM-LSS}

 The XMM-LSS provides a source density of $\sim 300$ /\dd\ down to a flux limit of 4$~10^{-15}$ \flux\ in the [0.5-2] keV band (95\% level completeness limit). This constitutes the largest deep AGN X-ray sample over a single field. The special  relevance of the sample also comes from its unique multi-$\lambda$ coverage.  
 \subsection{The spatial distribution}
The XMM-LSS data set allowed, for the first time, the study of the angular distribution of   faint AGNs over an area of 5 \dd. We found a significant clustering in the soft band and none in the hard band.  A sub-sample of $\sim 200$ sources with hard X-ray count ratios,  likely  dominated by obscured AGNs, does show a positive signal allowing for a large angular correlation length at the  $3\sigma$ level (Gandhi et al 2006).
 \subsection{Spectral properties}
A dedicated study of some 100 AGN selected in the hard band reveals a mismatch   between the classification based on the characteristics of the optical emission lines and the classification given by the X-ray spectroscopy. This led  to question some aspects of the AGN unified scheme (Garcet et al 2007). The many CFHTLS and SIWRE flux data points allowed us to perform SED fits on the XMM-LSS point source population. From this, it was possible to classify the AGN (star formation, type 1 or 2 AGN, Seyfert) and to obtain photometric redshifts. Combining with the X-ray spectral data points, we demonstrate that the SED properties are continuous through the various classes (Tajer et al 2007, Polletta et al 2007).

 \section{Data releases}
 Cluster data (positions, redshifts, L$_{X}$, T$_{X}$, mass estimates) are published for the first 5 \dd\  along with the scientific analysis (Valtchanov et al 2004 , Willis et al 2005, P07).
 The complete source catalogue with optical data and thumb nail images is also public (Pierre et al 2007).
 Data can be retrieved from the CDS or, in a more extensive form, via the consortium data bases for the cluster\footnote{http://l3sdb.in2p3.fr:8080/l3sdb/} and source\footnote{http://cosmos.iasf-milano.inaf.it/~lssadmin/Website/LSS/Query/} catalogues.

 \section{Next decade with XMM: A 200 \dd\ wide survey}   
 
\begin{figure}
\includegraphics[width= 8 cm]{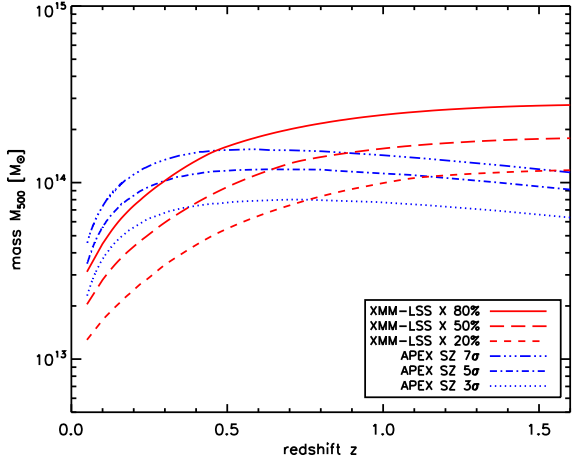}
 \vspace{-0.3cm}
 \caption{{\bf Comparison between the XMM-LSS and Sunyaev-Zel'dovich sensitivities} in 
terms of limiting mass. The red lines show various {\em measured} detection probability thresholds for the C1 clusters. The blue lines are the {\em predictions} for the 
10$\mu$K APEX survey, currently observing the XMM-LSS field. For the regime of interest ($z<1$), the X-ray observations  are at least as efficient as the S-Z ones in terms of cluster detection.}
\label{SZ}
\end{figure}

 During the past years, we  explored a number of issues regarding  cluster detection and science with the XMM-LSS survey. In addition to the C1 clusters, a complementary cluster sample of about the same size has been identified in the XMM-LSS, but for which the selection criteria are less well defined (Adami et al in prep). We have also detected a number of $z>1$ clusters (Andreon et al 2005, Bremer et al 2006).  The determination of the luminosity, temperature and mass (hydrostatic hypothesis) of the C1clusters from survey data appears to provide quantities reliable for statistical cosmological studies\footnote{The mass of XLSS 29 at $z = 1.05$ was measured to be $1.4~10^{14} ~M_{\odot}$ with survey data (P07) and  $1.8 \pm 0.5~10^{14} ~M_{\odot}$ from a subsequent 80 ks XMM pointing (Maughan et al 2007)}. A weak lensing analysis over part of the XMM-LSS area shows promising prospects for constraining independently the slope and the normalisation of the M-T relation (Berge et al 2007).  A further step in constraining cluster masses  is to be reached by the upcoming generation of Sunyaev- Zel'dovich surveys which, in principle, with a sensitivity of $10 \mu K$, are  well matched to mass range of the XMM-LSS survey (Fig. \ref{SZ}).  
 
Further, we have shown that, in order to increase the  precision  of the L-T relation evolution, it is more efficient to increase the cluster sample than the accuracy on the temperature measurements -  a useful tip for a proper use of  future XMM observing time. This is due to the large intrinsic dispersion of the L-T relation itself (P07).

\begin{figure}
\includegraphics[width= 7 cm]{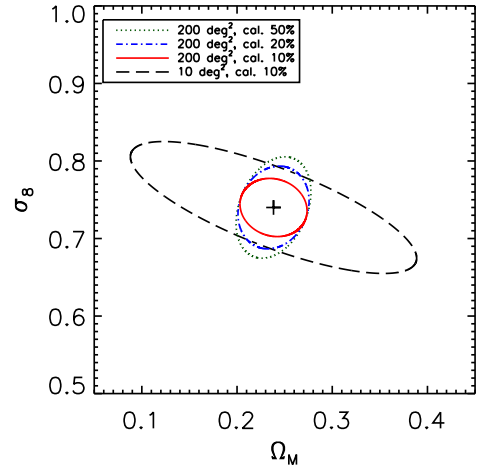}
  \vspace{-0.3cm}
\includegraphics[width= 7 cm]{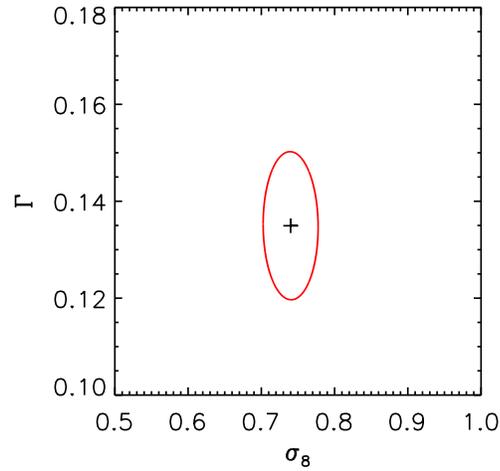}
\caption{ {\bf Top: Constraints
on the \boldmath$\sigma_{8}-\Omega_{m}$ plane from $dn/dz$ +  $\xi(r)$.}
 All contours are 1$\sigma$ for C1 clusters,
marginalised over $\Omega_{\Lambda}$. ({\sl dots}) The C1 X-ray cluster
population alone (6/\dd) over 200 \dd; adding information from
 S-Z ({\sl dash-dot})  and   weak lensing ({\sl solid}) mass measurements;
({\sl dash})
same as solid but for 10 \dd\ coverage. X-ray masses are taken to be
accurate to 50\%, adding S-Z then weak lensing  data reduces this to  20\% then to
10\%; the latter giving  an accuracy on $\sigma_8$ of
6 \%. {\bf Bottom: Constraints on the \boldmath$\sigma_{8}-\Gamma$
plane from $\xi(r)$ + $dn/dz$.} Contours are 1 $\sigma$ for the C1
population, 200 \dd, mass accuracy of 10\% and marginalisation over
$\Omega_m$ and $\Omega_{\Lambda}$.  The expected accuracy on $\Gamma$ is 10\%.}
\label{merite}
\end{figure}

In this way, a 200 \dd\ survey with 10 ks XMM pointings\footnote{such a survey would require some 24 Ms with the current observing settings, which could be decreased by about 1/3 when the foreseen mosaicing mode with a reduced pn overhead  is implemented}, would allow  definitively discriminating not only between cluster self-similar evolution and no evolution, but also between  other theoretically justified models based on non-gravitational physics (Fig. \ref{LT}). Such a 200 \dd\ survey would moreover overcome sample variance problems and provide more than a thousand C1 clusters as well as some 100 clusters at $z>1$. Having determined the cluster evolution rate, an important degeneracy would be removed in the cosmological interpretation of the cluster number counts (Fig. \ref{dndz}) leaving a handle on the equation of state of the Dark Energy (P07). The cluster sky distribution  provides additional constraints on the cosmological parameters and a 200 \dd\ survey will not only allow a self-sufficient and accurate determination of    $\sigma_{8}$, i.e. 5 \%, but also of  the slope of the matter power spectrum, $\Gamma$, i.e. 10\% (Fig. \ref{merite}).

  \end{document}